\documentclass[12pt]{article}
\usepackage{graphicx}

\def \Bbar{\bar B}

\def \fbar{\bar{f}}

\def \b{{\cal B}}

\def \babar{{\sc BaBar}}
\def \bea{\begin{eqnarray}}
\def \beq{\begin{equation}}

\def \eea{\end{eqnarray}}
\def \eeq{\end{equation}}

\def \({\left(}
\def \){\right)}
\def \[{\left[}
\def \]{\right]}
\def \nn{\nonumber}

\def \s{\sqrt{2}}

\def \ga{\gamma}
\def \Ga{\Gamma}
\def \de{\delta}

\def \bea{\begin{eqnarray}}
\def \beq{\begin{equation}}
\def \eea{\end{eqnarray}}
\def \eeq{\end{equation}}
\def \nl{\nn\\}

\def \cA{{\cal A}}
\def \ocA{\overline{\cA}}
\def \cB{{\cal B}}
\def \cT{{\cal T}}
\def \cP{{\cal P}}
\begin{document}
\rightline{TECHNION-PH-13-5}
\rightline{EFI 13-4}
\rightline{UdeM-GPP-TH-13-223}
\rightline{June 2013}
\bigskip
\centerline{\bf CP ASYMMETRIES IN THREE-BODY $B^\pm$ DECAYS}
\centerline{\bf TO CHARGED PIONS AND KAONS}
\bigskip
\centerline{Bhubanjyoti Bhattacharya}
\centerline{\it Physique des Particules, Universit\'e de Montr\'eal}
\centerline{\it C.P. 6128, succ.\ centre-ville, Montr\'eal, QC, Canada H3C 3J7}
\medskip

\centerline{Michael Gronau}
\centerline{\it Physics Department, Technion -- Israel Institute of Technology}
\centerline{\it Haifa 3200, Israel}
\medskip

\centerline{Jonathan L. Rosner}
\centerline{\it Enrico Fermi Institute and Department of Physics}
\centerline{\it University of Chicago, 5620 S. Ellis Avenue, Chicago, IL 60637}
\bigskip
\begin{quote}
CP asymmetries have been measured recently by the LHCb collaboration in
three-body $B^+$ decays to final states involving charged pions and kaons.
Large asymmetries with opposite signs at a level of about $60\%$ have been
observed in $B^\pm\to \pi^\pm({\rm or}~K^\pm)\pi^+\pi^-$ and $B^\pm \to 
\pi^\pm K^+K^-$ for restricted regions in the Dalitz plots involving $\pi^+\pi^-$ 
and $K^+K^-$ with low invariant mass. U-spin is shown to predict 
corresponding $\Delta S=0$ and $\Delta S=1$ asymmetries with opposite signs 
and inversely proportional to their branching ratios, 
in analogy with a successful relation predicted thirteen years ago between asymmetries 
in $B_s\to K^-\pi^+$ and $B^0 \to K^+ \pi^-$.  We compare these predictions 
with the measured integrated asymmetries.  Effects of specific resonant or
non-resonant partial waves on enhanced asymmetries for low-pair-mass regions
of the Dalitz plot are studied in $B^\pm \to \pi^\pm \pi^+\pi^-$.
The closure of low-mass $\pi^+\pi^-$ and $K^+K^-$ 
channels involving only $\pi\pi \leftrightarrow K\bar K$ rescattering may explain 
by CPT approximately equal magnitudes and opposite signs measured in 
$B^\pm\to \pi^\pm\pi^+\pi^-$ and $B^\pm \to \pi^\pm K^+K^-$.
\end{quote}

\leftline{PACS numbers: 13.25.Ft, 13.25.Hw, 14.40.Lb, 14.40.Nd}
\bigskip

\section{Introduction}

CP asymmetries in two-body and quasi-two-body charmless $B$ and $B_s$
decays have played an important role in testing the Cabibbo-Kobayashi-Maskawa
(CKM) framework and, with high precision, may probe virtual effects of heavy
new particles.  Isospin symmetry, which holds well in strong interactions, has
been applied experimentally in two well-known cases to CP asymmetries in
strangeness-conserving $B\to \pi\pi, \rho\rho$ and strangeness-changing $B\to
K \pi$ decays.

In the first example~\cite{Gronau:1990ka}, CP asymmetries measured in $B^0 \to
\pi^+\pi^-$ and $B^0\to \rho^+\rho^-$ provide the most precise source of
information for determining $\alpha\equiv {\rm Arg}(-V^*_{tb}V_{td}/V^*_{ub}
V_{ud})$~\cite{CKMfitter}.  Two inputs of this method, the direct asymmetries
$A_{CP}(B^0\to \pi^+\pi^-)$ and $A_{CP}(B^0 \to \rho^+\rho^-)$ due to
interference of tree and penguin amplitudes, may be affected by new heavy
particles entering the $b \to d$ penguin loop. This could show up through an
inconsistency between this measurement of $\alpha$ and a future improved
determination of $\gamma\equiv {\rm Arg }(-V^*_{ub}V_{ud}/V^*_{cb}V_{cd})$ in
$B\to D^{(*)}K^{(*)}$~\cite{GLW,HFAG}.  In the second example an isospin sum
rule has been proposed~\cite{Gronau:2005kz} combining CP rate differences in
all four $B\to K\pi$ decays, $B^0\to K^+\pi^-, K^0\pi^0$ and $ B^+ \to
K^+\pi^0, K^0\pi^+$.  A violation of this sum rule, rather than
merely the nonzero difference observed between CP asymmetries
in $B^0 \to K^+\pi^-$ and $B^+\to K^+\pi^0$~\cite{HFAG} (often named ``the
$K\pi$ puzzle"~\cite{Lin:2008zzaa,Gronau:1998ep,QCD}), would be 
unambiguous evidence for new
physics in $b \to s q \bar q$ transitions. This test requires a substantial
improvement in the measurement of $A_{CP}(B^0 \to K^0\pi^0)$~\cite{HFAG}.

An application of isospin symmetry to three-body $B$ decays of pions and kaons
is too involved for studying CP asymmetries in these decays because {\em five}
independent isospin amplitudes are needed for describing merely the subset of
decays to three kaons~\cite{Grossman:2003qp,Gronau:2003ep}. In contrast, U-spin
symmetry, an SU(2) subgroup of flavor SU(3) under which the pairs of quarks
$(d,s)$ and mesons $(\pi^-,K^-)$ transform like doublets, seems potentially
powerful for studying asymmetries in all three-body charged $B$ decays
involving charged pions and kaons, $B^+ \to K^+ \pi^+\pi^-, K^+ K^+K^-$,
$\pi^+\pi^+\pi^-,  \pi^+K^+K^-$.  All four processes involve only {\em two}
independent U-spin amplitudes~\cite{Gronau:2003ep}.
One should be aware of possible U-spin breaking effects of order $30\%$,
expected  in hadronic amplitudes  and consequently in  CP asymmetry relations.

Dalitz-plot analyses of these three-body processes have been carried out by
the \babar~and Belle collaborations for 
$B^+ \to K^+ \pi^+\pi^-$ \cite{Aubert:2008bj,Garmash:2005rv},
$B^+ \to K^+ K^+K^-$ \cite{Aubert:2006nu,Lees:2012kxa,Garmash:2004wa,%
Lees:2013kua}, $B^+ \to \pi^+\pi^+\pi^-$ 
\cite{Aubert:2005sk,Aubert:2009av}, and $B^+ \to \pi^+K^+K^-$ \cite{Aubert:2007xb}.
Three-body charmless $B$ and $B_s$ decays have been  studied under various 
assumptions in Ref.\ \cite{Th-B3P}.

The LHCb collaboration has recently reported measurements of CP asymmetries for
all four three-body $B^+$ decay modes involving charged pions and kaons. The
measurements include two asymmetries in decays to strangeness-one final
states~\cite{Aaij:2013sfa},
\bea\label{S=1}
A_{CP}(B^+ \to K^+ \pi^+\pi^-) & = & +0.032 \pm 0.008({\rm stat}) \pm 0.004({\rm syst})
\pm 0.007(J/\psi K^+)~,\nonumber \\
A_{CP}(B^+ \to K^+ K^+K^-) & = & -0.043 \pm 0.009({\rm stat}) \pm 0.003 ({\rm syst})
\pm 0.007(J/\psi K^+)~,
\eea
with significance of $2.8\sigma$ and $3.7\sigma$.
A very recent \babar~result~\cite{Lees:2013kua}, $ A_{CP}(B^+ \to K^+
K^+K^-) = -0.017^{+0.019}_{-0.014} \pm 0.014$, is consistent with the LHCb 
measurement within $1.1 \sigma$.
 
Two other asymmetries have been measured by LHCb in decays to strangeness-zero
states~\cite{LHCb3P0},
\bea\label{S=0}
A_{CP}(B^+ \to \pi^+\pi^+\pi^-) & = & +0.120 \pm 0.020({\rm stat}) \pm
0.019({\rm syst}) \pm 0.007(J/\psi K^+)~,\nonumber \\
A_{CP}(B^+ \to \pi^+K^+K^-) & = & -0.153 \pm 0.046({\rm stat}) \pm
0.019({\rm syst}) \pm 0.007(J/\psi K^+)~,
\eea
with significance of $4.2\sigma$ and $3.0\sigma$.

Considerably larger CP asymmetries with same signs as the above were measured
in the latter two decay modes for localized regions of phase space. The two
regions, corresponding to pairs of $\pi^+\pi^-$ and $K^+K^-$ with low invariant
mass, $m^2_{\pi^+\pi^-\,{\rm low}} <  0.4~{\rm GeV}^2/c^4$ (the other pion pair
obeying $m^2_{\pi^+\pi^-\,{\rm high}}$ $> 15~{\rm GeV}^2/c^4$) and
$m^2_{K^+K^-} < 1.5~{\rm GeV}^2/c^4$, involve the following asymmetries
\cite{LHCb3P0}:
\bea\label{local}
A_{CP}(B^+ \to \pi^+(\pi^+\pi^-)_{{\rm low}\,m}) & = & +0.622 \pm 0.075 \pm
0.032  \pm 0.007~,\nonumber \\ A_{CP}(B^+ \to \pi^+(K^+K^-)_{{\rm low}\,m})
& = & -0.671 \pm 0.067 \pm 0.028 \pm 0.007~.
\eea
Enhancements have also been observed in $\Delta S=1$ asymmetries (\ref{S=1})
for low-mass $\pi^+\pi^-$ and $K^+K^-$ 
pairs, $0.08~{\rm GeV}^2/c^4 < m^2_{\pi^+\pi^-} <
0.66~{\rm GeV}^2/c^4$, $m^2_{K^+\pi^-} <15~{\rm GeV}^2/c^4$ and 
$1.2~{\rm GeV}^2/c^4 < m^2_{K^+K^-\,{\rm low}} < 2.0~{\rm GeV}^2/c^4$, 
$m^2_{K^+K^-\,{\rm high}} < 15~{\rm GeV}^2/c^4$~\cite{Aaij:2013sfa}:  
\bea\label{local1}
A_{CP}(B^+ \to K^+(\pi^+\pi^-)_{{\rm low}\,m}) & = & +0.678 \pm 0.078 \pm
0.032  \pm 0.007~,\nonumber \\ A_{CP}(B^+ \to K^+(K^+K^-)_{{\rm low}\,m})
& = & -0.226 \pm 0.020 \pm 0.004 \pm 0.007~.
\eea

The purpose of this Letter is to study these measured asymmetries theoretically,
trying to understand their pattern within the CKM framework.  While the
inclusive asymmetries (\ref{S=1}) and (\ref{S=0}) require integration over an
entire three-body phase space, the asymmetries (\ref{local}) and (\ref{local1}) 
for localized regions of phase space may depend on resonance
behavior dominating these regions. Thus different methods may have to be
applied for analyzing total and localized asymmetries.

In Section \ref{sec:Uspin} we show that total CP rate differences for pairs of
$\Delta S=0$ and $\Delta S=1$ three-body $B^+$ decay processes, where final
states are related to each other by a U-spin reflection, are equal in
magnitudes and have opposite signs.  We reiterate a general proof presented
in Ref.~\cite{Gronau:2000zy}, aimed mainly at pairs of two-body and quasi-two-body
$B$ and $B_s$ decays (see also~\cite{He:1998rq,Fleischer:1999pa,Gronau:2000md}), 
mentioning only briefly the pair $B^+ \to K^+ \pi^+\pi^-$ and $B^ + \to \pi^+ K^+K^-$.
(Ref.\,\cite{Gronau:2003ep} mentioned briefly a similar relation for $B^+\to \pi^+\pi^+\pi^-$ 
and $B^+ \to K^+K^+K^-$.)
Section \ref{sec:Local} studies sources of enhanced CP asymmetries for localized 
regions of phase space involving low-mass $\pi^+\pi^-$ and $K^+K^-$ pairs, 
pointing out a possible approximate asymmetry relation following from CPT, 
while Section \ref{sec:Conc} concludes.

\section{U-spin relates $\Delta S=0, 1$ CP rate asymmetries\label{sec:Uspin}}
The low-energy effective weak Hamiltonian describing $\Delta S =1$ $B$ decays
is~\cite{Buchalla:1995vs}
\beq\label{Heff1}
{\cal H}^{\Delta S=1}_{\rm eff} = \frac{G_F}{\s}\left[V^*_{ub}V_{us}\left(\sum^2_1 C_i(\mu)
Q^{us}_i +\sum^{10}_3 C_i(\mu) Q^s_i\right ) + V^*_{cb}V_{cs}\left(\sum^2_1 C_i(\mu)
Q^{cs}_i +\sum^{10}_3 C_i(\mu) Q^s_i\right )\right]~,
\eeq
where $C_i(\mu)$ are scale-dependent Wilson coefficients and $Q^{qs}_i (q=u, c), Q^q_i$
are four-quark operators. Suppressing chiral and color structure of these operators and
denoting quark charges by $e_q$, the flavor structure of the operators is given by
\bea
 Q^{qs}_{1,2} & = & \bar b q\bar q s~,~q=u, c~,\nonumber\\
 Q^s_{3,..,6} & = & \bar b s \sum_{q'=u,d,s,c} \bar q' q'~,\nonumber\\
 Q^s_{7,..,10} & = & \frac{3}{2} \bar b s \sum_{q'=u,d,s,c} e_{q'}\bar q' q'~.
\eea

Focusing on U-spin properties of these twelve operators we note that, since  $u, c, b$
and $\bar d d + \bar s s$ are U-spin singlets, each of these operators
represents an $s$ (``down") component of a U-spin doublet operator, so that
one can write in short
\beq\label{Us}
{\cal H}^{\Delta S=1}_{\rm eff} = V^*_{ub}V_{us}\,U^s + V^*_{cb}V_{cs}\,C^s~,
\eeq
where $U$ and $C$ are two independent U-spin doublet operators. Similarly, the
$\Delta S =0$ effective Hamiltonian ${\cal H}^{\Delta S =0}_{\rm eff}$, in which
one replaces $s$ by $d$ in Eq.~(\ref{Heff1}), involves $d$ (``up") components
of the same two U-spin doublet operators $U$ and $C$ multiplying different CKM
factors, $V^*_{ub}V_{ud}$ and $V^*_{cb}V_{cd}$,
\beq\label{Ud}
{\cal H}^{\Delta S=0}_{\rm eff} = V^*_{ub}V_{ud}\,U^d + V^*_{cb}V_{cd}\,C^d~.
\eeq

A very simple implication of Eqs.\ (\ref{Us}) and (\ref{Ud}) is obtained by
comparing two decay processes,  $\Delta S=1$ and $\Delta S=0$, in which initial
and final states are each other's U-spin reflections,
\beq
U_r: d \leftrightarrow s~.
\eeq
We write the $\Delta S=1$ amplitude for a generic process $B \to f$ in the form
\beq\label{s}
A(B\to f,~\Delta S =1) = V^*_{ub}V_{us}\,A_u + V^*_{cb}V_{cs}\,A_c~,
\eeq
where $A_u\equiv \langle f|U^s|B\rangle$ and $A_c \equiv \langle f|C^s|B\rangle$
are complex amplitudes involving CP-conserving phases.
The  $\Delta S =0$ amplitude for the corresponding U-spin reflected process,
$U_rB \to U_rf$, is then given by
\beq\label{d}
A(U_rB\to U_rf,~\Delta S =0) = V^*_{ub}V_{ud}\,A_u + V^*_{cb}V_{cd}\,A_c~,
\eeq
where we used $\langle U_r f|U^d|U_r B\rangle = \langle f|U^s|B\rangle \equiv A_u,
\langle U_r f|C^d|U_r B\rangle = \langle f|C^s|B\rangle \equiv A_c$.

In the case of three-body $B^+$ decays (where $B^+$ is invariant under $U_r$)
the amplitudes in (\ref{s}) and (\ref{d}) depend
on the same corresponding final particle momenta. For instance
\bea
A(B^+ \to K^+(p_1)\pi^+(p_2)\pi^-(p_3)) & = &  V^*_{ub}V_{us}\,A_u(p_1,p_2,p_3) +
V^*_{cb}V_{cs}\,A_c(p_1,p_2,p_3)~,\nonumber \\
A(B^+ \to \pi^+(p_1)K^+(p_2)K^-(p_3) & = & V^*_{ub}V_{ud}\,A_u(p_1,p_2,p_3) +
V^*_{cb}V_{cd}\,A_c(p_1,p_2,p_3)~.
\eea
 
Applying CP-conjugation to (\ref{s}) and (\ref{d}), one has
 \bea\label{sdbar}
A(\Bbar \to \fbar,~\Delta S =-1) & = & V_{ub}V^*_{us}\, \bar A_u + V_{cb}V^*_{cs}\, \bar A_c~,
\nonumber\\
A(U_r\Bbar\to U_r\fbar,~\Delta S =0) & = & V_{ub}V^*_{ud}\,\bar A_u + V_{cb}V^*_{cd}\,\bar A_c~.
\eea
Here $\bar A_{u, c} = A_{u, c}$ for two-body decays and $\bar A_{u, c} = 
A_{u, c}(-\vec p_1, -\vec p_2, -\vec p_3)$ for three-body decays. 
Thus
\beq\label{asymB}
\vert A(B \to f)\vert^2 - \vert A(\Bbar \to \fbar)\vert^2 =
2{\rm~Im}(V^*_{ub}V_{us}V_{cb}V^*_{cs}){\rm Im}(A^*_uA_c + \bar A^*_u \bar A_c)~,
\eeq
while
\beq
\vert A(UB \to Uf)\vert ^2 - \vert A(U\Bbar \to U\fbar)\vert^2 =
2{\rm~Im}(V^*_{ub}V_{ud}V_{cb}V^*_{cd}){\rm Im}(A^*_u A_c + \bar A^*_u \bar A_c)~.
\eeq
Unitarity of the CKM matrix implies~\cite{Jarlskog:1985ht}
\beq
{\rm Im}(V^*_{ub}V_{us}V_{cb}V^*_{cs}) =
- {\rm Im}(V^*_{ub}V_{ud}V_{cb}V^*_{cd})~,
\eeq
leading to a general U-spin relation~\cite{Gronau:2000zy}
\beq\label{asym}
\vert A(B \to f)\vert^2 - \vert A(\Bbar \to \fbar)\vert^2 =
-[\vert A(UB \to Uf)\vert ^2 - \vert A(U\Bbar \to U\fbar)\vert^2]~.
\eeq

In the case of three-body $B^+$ decays one integrates this momentum-dependent
amplitude relation over three-body phase space to obtain a corresponding relation
between CP rate differences.
Denoting rates by the final charged particles,
the following relations are expected to hold in the  U-spin symmetry limit:
\bea
\Gamma(\pi^- K^- K^+) - \Gamma(\pi^+ K^+ K^-) & = &
- [\Gamma(K^-\pi^-\pi^+) -\Gamma(K^+ \pi^+\pi^-)]~,\nonumber\\
\Gamma(\pi^-\pi^- \pi^+) - \Gamma(\pi^+\pi^+\pi^-)] & = &
- [\Gamma(K^-K^-K^+) - \Gamma(K^+ K^+ K^-)]~,
\eea
or
\bea\label{ratio1}
\frac{A_{CP}(B^+\to \pi^+ K^+ K^-)}{A_{CP}(B^+ \to K^+ \pi^+ \pi^-)} & = &
-\frac{\b(B^+\to K^+ \pi^+\pi^-)}{\b(B^+ \to \pi^+ K^+ K^-)}~,\\
\frac{A_{CP}(B^+ \to \pi^+ \pi^+\pi^-)}{A_{CP}(B^+ \to K^+K^+K^-)} & = &
-\frac{\b(B^+ \to K^+K^+ K^-)}{\b(B^+ \to \pi^+ \pi^+ \pi^-)}~.
\label{ratio2}
\eea

Branching fractions for the four three body $B^+$ decay modes involving
charged pions and kaons are given in Table \ref{tab:B}~\cite{HFAG}.
%
\begin{table}
\caption{Branching fractions for three-body $B^+$ decays to charged pions and
kaons~\cite{HFAG}.
\label{tab:B}}
\begin{center}
\begin{tabular}{c c} \hline \hline
Final state & Branching fraction ($10^{-6}$)\\ \hline
 $K^+\pi^+\pi^-$  & $51.0 \pm 3.0$ \\
 $K^+K^+K^-$ & $34.0 \pm 1.0$  \\
 $\pi^+\pi^+\pi^-$ & $15.2 \pm 1.4$  \\
 $\pi^+K^+K^-$ & $5.0 \pm 0.7$ \\
\hline \hline
\end{tabular}
\end{center}
\end{table}
Table \ref{tab:Asym-ratios} compares U-spin symmetry predictions for ratios of
asymmetries using Eqs.~(\ref{ratio1}) and (\ref{ratio2}) with the LHCb results
(\ref{S=1}) and in (\ref{S=0}).  U-spin predicts the two ratios of $\Delta S=0$
and $\Delta S=1$ asymmetries to be negative, as measured, and larger than one
--- inversely proportional to corresponding branching ratios.
%
\begin{table}
\caption{U-spin predictions for asymmetry ratios (\ref{ratio1}) and
(\ref{ratio2}) compared with LHCb measurements (\ref{S=1}) and (\ref{S=0}).
\label{tab:Asym-ratios}}
\begin{center}
\begin{tabular}{c c c} \hline \hline
Asymmetry ratio & U-spin prediction & LHCb result\\ \hline
$A_{CP}(B^+\to \pi^+ K^+ K^-)/A_{CP}(B^+ \to K^+ \pi^+ \pi^-)$ &
$-10.2 \pm 1.5$ & $-4.8 \pm 2.3$ \\
$A_{CP}(B^+ \to \pi^+ \pi^+\pi^-)/A_{CP}(B^+ \to K^+K^+K^-)$ &
$-2.2 \pm 0.2$ & $-2.8 \pm 1.0$ \\
\hline \hline
\end{tabular}
\end{center}
\end{table}

The modest violation of U-spin seen in the top line of Table
\ref{tab:Asym-ratios} 
(currently at $2.0\sigma$) is not surprising given the very different
resonant substructure of the two Dalitz plots.  Such sources of U-spin
breaking are absent in the successful prediction \cite{Gronau:2000md} of
the large negative asymmetry ratio in $B_s\to K^-\pi^+$ and
$B^0\to K^+\pi^-$~\cite{HFAG},
\beq\label{BKpitoBsKpi}
\frac{A_{CP}(B_s \to K^- \pi^+)}{A_{CP}(B^0 \to K^+ \pi^-)}
 = -\frac{\tau(B_s)}{\tau(B^0)}
\frac{\b(B^0\to K^+\pi^-)}{\b(B_s\to K^-\pi^+)} = - 3.6 \pm 0.4~.
\eeq
This prediction should be compared with the world-averaged
asymmetries~\cite{HFAG} recently updated by LHCb
measurements~\cite{Aaij:2013iua},
\beq
A_{CP}(B^0 \to K^+\pi^-) = -0.082 \pm 0.006~,~~~~A_{CP}(B_s \to K^- \pi^+) =
0.26 \pm 0.04~,
\eeq
implying $A_{CP}(B_s\to K^-\pi^+)/A_{CP}(B^0\to K^+\pi^-) = - 3.2 \pm
0.5$ in good agreement with (\ref{BKpitoBsKpi}).
The above prediction is subject to first order U-spin breaking
\cite{Gronau:1995hm,Nagashima:2007qn,Imbeault:2011jz,He:2013vta,Gronau:2013mda} 
in the sum
\beq 
\frac{A_{CP}(B_s \to K^- \pi^+)}{A_{CP}(B^0 \to K^+ \pi^-)} + 
\frac{\tau(B_s)}{\tau(B^0)}\frac{\b(B^0\to K^+\pi^-)}{\b(B_s\to K^-\pi^+)} = 0.4 \pm 0.6~,
\eeq
which should be compared to each of the two terms in this sum whose magnitudes are 
each around 3 - 4.
  
\section{Asymmetries for low mass $\pi^+\pi^-$ and $K^+K^-$ pairs
\label{sec:Local}}

The strangeness-conserving CP asymmetries exhibited in Eq.\ (\ref{local}) have
two distinguishing features.  (1) They are large, very close to maximal.  The
relative weak phase of the $\bar b \to \bar d$ penguin and the $\bar b \to \bar
u u \bar d$ tree amplitude is $\gamma$, whose sine is very large
\cite{CKMfitter}.  (2) They are opposite in sign.
The U-spin relations discussed in the previous Section would have then implied 
smaller asymmetries, but also of opposite signs, for restricted regions of phase space
in the $|\Delta S| = 1$ transitions $B^+ \to K^+ \pi^+ \pi^-$ and $B^+ \to K^+
K^+ K^-$.  While the opposite sign relation holds between Eqs.\ (\ref{local}) and 
(\ref{local1}), the large value of $A_{CP}(B^+\to K^+(\pi^+\pi^-)_{{\rm low}\,m})$ 
exhibits sizeable U-spin breaking relative to $A_{CP}(B^+\to 
\pi^+(K^+K^-)_{{\rm low}\,m})$ due to different resonant structures of $\pi^+\pi^-$
and $K^+K^-$.  

The Dalitz plot for $B^\pm \to \pi^\pm \pi^+ \pi^-$ contains a prominent
$\rho^0$ band in the spectrum of the low-mass $\pi^+ \pi^-$ pair.  This
band would be equally populated near both ends in the absence of interference
with other partial waves.  However, the extremity with high $m^2_{\pi^+ \pi^-
{\rm~high}}$ is visibly depopulated in comparison with the extremity with
low $m^2_{\pi^+ \pi^- {\rm~high}}$, strongly suggesting interference with a
strong S-wave amplitude.  This feature has led to the proposal that such
interference is responsible for the pronounced CP asymmetry in the first Eq.\
(\ref{local}) \cite{Zhang:2013oqa}.  With a suitable relative strong phase,
this mechanism could explain both the overall asymmetry and the fact that it is
enhanced when selecting events with low $m^2_{\pi^+ \pi^- {\rm~low}}$ and high
$m^2_{\pi^+ \pi^- {\rm~high}}$.

An attempt to account for strong phases through resonant substates is
frustrated by incomplete information on the decays $B \to P S$, where $P$
and $S$ denote pseudoscalar and scalar mesons.  A recent fit to these decays
based on QCD factorization \cite{Cheng:2013fba} reaches different conclusions
depending on which scalar mesons are ascribed to a $^3P_0$ $q \bar q$ nonet
and which are labeled as tetraquarks or mesonic molecules.  Relative strong
phases {\it are} available in fits to $B \to PV$ decays, where $V$ denotes
a vector meson \cite{Chiang:2003pm}.

To show that a CP asymmetry for $B^\pm \to \pi^\pm \pi^+\pi^-$ as large as that
in Eq.\ (\ref{local}) is plausible, we supplement the Dalitz plot analysis of
Ref.\ \cite{Aubert:2009av} with an amplitude
corresponding to $f_0(500) \pi^\pm$, where $f_0(500)$ is an S-wave $I=0~\pi
\pi$ resonance assumed for present purposes to be a $q \bar q$ state.  We may
then relate a penguin amplitude for $B^\pm \to \pi^\pm f_0(500)$ via SU(3) to
the penguin amplitude assumed to dominate $B^\pm \to \pi^\pm K_0^*(1430)$.

\subsection{$B \to PV$}

Following Ref.\ \cite{Chiang:2003pm} we may write the relevant amplitudes for
$B^+\to\rho^0\pi^+$ as follows:
\bea\label{eq:rho}
\cA(B^+\to\rho^0\pi^+) &=& -\frac{1}{\s} \(t_V + c_P + p_V - p_P\)~, \nl
&=& - |\cP| - |\cT|~ e^{i(\de_V ~+~ \ga)}~.
\eea
The assumption in Ref.\ \cite{Chiang:2003pm} is that $p_V = - p_P$ and these
amplitudes are chosen to be real as shown in Fig.\ 2
in Ref.\ \cite{Chiang:2003pm}. Also $t_V$ and $c_P$ have 
zero relative phase, while the relative strong and weak phases of $t_V$ with 
respect to $p_V$ are $\delta_V \sim -18^\circ$ and $\ga \sim 65^\circ$. 
Under these conditions then:
\bea\label{PT}
|\cP| &=& \s |p_V| ~=~ \s |p_P| ~=~ \s \times 7.5~{\rm eV} ~=~ 1.06\times10^{-5}
~{\rm MeV}~, \nl
|\cT| &=& \frac{|t_V| + |c_P|}{\s} ~=~ \frac{1}{\s}(30.3+5.3)~{\rm eV} ~=~ 2.51
\times10^{-5}~{\rm MeV}~.
\eea
The CP-conjugate process will thus have the amplitude given as
\bea\label{eq:rhob}
\ocA(B^-\to\rho^0\pi^-) &=& - |\cP| - |\cT|~ e^{i(\de_V ~-~ \ga)}~.
\eea
In Ref.\ \cite{Chiang:2003pm} the strong phase of $p_P$ is taken to be zero and
(as a consequence of the assumption $p_V = - p_P$) the strong phase of $p_V$
is taken to be $\pi$. This is based on the relative P-wave amplitude between
the final-state particles \cite{Lipkin:1980tk}.  Following this reasoning, in
the $B \to PS$ case, we take 
$\tilde p_S = \tilde p_P$.

Using the above we estimate the CP-averaged amplitude and CP asymmetry for
the process as follows:
\bea
|\cA|^2_{\rm avg} &=& \frac{|\ocA|^2 + |\cA|^2}{2} \nl
&=& \(|\cP|^2 + |\cT|^2 + 2 |\cP| |\cT| \cos\de_V\cos\ga\) \nl
&=& \(30.9~{\rm eV}\)^2~~~~
~~~~~({\cal B}~\sim~9\times10^6~\cite{HFAG})~,\\ 
\label{ACPrho}
A_{\rm CP} &=& \frac{|\ocA|^2 - |\cA|^2}{|\ocA|^2 + |\cA|^2} \nl
&=& \frac{2 |\cP| |\cT|\sin\de_V\sin\ga}{|\cA|^2_{\rm avg}} \nl
&=& -0.16~.
\eea
The CP asymmetry is very sensitive to changes in the strong phase $\de_V$.
Its value in 2004, to which $\de_V$ was fitted in Ref.\ \cite{Chiang:2003pm},
was $-0.19 \pm 0.11$. 
While its slightly different current value~\cite{HFAG} $0.18^{+0.09}_{-0.17}$
is not significantly inconsistent with (\ref{ACPrho}), the central value favors 
$\delta_V \sim +18^\circ$.

\subsection{$B \to PS$}

We consider $B\to PS$ decays with some inputs from Ref.\ \cite{Cheng:2013fba}.
The decay $B^+\to K^*_0(1430)\pi^+$ is a
pure penguin $|\Delta S| = 1$ process, and its amplitude may be written as
\bea
\cA(B^+\to K^*_0(1430)\pi^+) ~=~ \tilde p'_P~.
\eea
We use the branching fraction quoted in Ref.\ \cite{Cheng:2013fba} (average
of Belle and BaBar): $\cB(B^+\to K^*_0(1430)\pi^+) = 45.1\times10^{-6}$,
finding  $|\tilde p'_P| = 7.2\times 10^{-5}$ MeV. The corresponding contribution
to a $\Delta S = 0$ process is then $\tilde p_P = \lambda |\tilde p'_P| = 1.66 \times
10^{-5}$ MeV, where we have used $\lambda \equiv \tan\theta_C = 0.23$.
   
Since $f_0(500)$ is a singlet, we assume that it is $\frac{1}{\s}(u{\bar u}
+ d{\bar d})$. One can then write the amplitude representation for 
$B^+\to f_0(500)\pi^+$ as follows:
\bea
\cA(B^+\to f_0(500)\pi^+) &=& - \frac{1}{\s}(\tilde p_S + \tilde p_P - \tilde t_S - 
\tilde c_P)~, \nl
&=& - \s~|\tilde p_P|~e^{i\de_{f_0}} + |\tilde T|~e^{i(\de_S + \ga)}~,
\eea
where we have assumed $\tilde p_S = \tilde p_P$.  We have defined $\tilde t_S +
\tilde c_P \equiv \s\,|\tilde T|e^{i(\delta_S + \gamma)}$, while $\de_{f_0}$ is
the relative strong phase between the penguin contributions in the amplitudes
for $\rho^0$ and $f_0$ modes.  $f_0(500)$ is a wide scalar resonance with
mass close to 500 MeV and width close to 540 MeV.  In the absence of any
reliable estimate for a tree contribution
in $B^+\to f_0(500) \pi^+$, we assume that the amplitude for this
process is penguin-dominated.  A tree contribution in this process would 
lead to a 3-body asymmetry by interference with $\cP$ which is seen in (25) 
to be suppressed relative to $\cT$.

We may then predict the amplitude and branching fraction for the process:
\bea\label{eq:f0}
\cA(B^+\to f_0(500)\pi^+) &~\sim& - \s~ |\tilde p_P|~e^{i\de_{f_0}}~, \\
\label{p_P}
|\cA(B^+\to f_0(500)\pi^+)| ~\sim~ \s~|\tilde p_P| &=& 2.35\times10^{-5}~{\rm MeV}~,\\
\label{f0pi+}
{\cal B}(B^+\to f_0(500)\pi^+) &\sim& 5.1\times10^{-6}~.
\eea
Our estimate (\ref{f0pi+}) is consistent with the current $90\%$ confidence level 
upper limit, ${\cal B}(B^+ \to f_0(500)\,\pi^+, f_0 \to \pi^+\pi^-) < 4.1\times 10^{-6}$ 
\cite{Aubert:2005sk}.
We note the comparable magnitudes of amplitudes $\cT$ and $\s\,\tilde p_P$ in
(\ref{PT}) and (\ref{p_P}) which may lead by their interference to a very large
asymmetry in  $B^+\to\pi^+\pi^-\pi^+$ for low-mass $\pi^+\pi^-$ pairs.

\subsection{$B^\pm\to\pi^\pm\pi^+\pi^-$ Dalitz plot}

We will now use an isobar analysis to reproduce the $B^\pm\to\pi^\pm\pi^+\pi^-$
Dalitz plot for low-mass $\pi^+\pi^-$ pairs.
In the isobar model with only two contributing components we can
write the amplitude for the three-body process in terms of the constituent
two-body processes as follows:
\bea\label{isobar}
\cA_{B^+\to\pi^+\pi^+\pi^-}(m^2_{\rm low}, m^2_{\rm high}) ~=~
c_\rho~F_\rho(m^2_{\rm low},
m^2_{\rm high}) + c_{f_0}~F_{f_0}(m^2_{\rm low}, m^2_{\rm high})~,
\eea
where the $c$'s are complex isobar coefficients and $F$'s are dynamical
functions of momenta $m^2_{low}$, and $m^2_{high}$ which respectively represent
the higher and lower invariant masses of the two $\pi^+\pi^-$ pairs in the
final state.  Since $f_0(500)$ is a scalar we use a simple Breit-Wigner form
for $F_{f_0}$ with $m_{f_0} = 500$ MeV and $\Ga_{f_0} = 540$ MeV. In case of
the $\rho^0$ it is standard to use a more specific Gounaris-Sakurai form, as
seen in Ref.\ \cite{Aubert:2009av}. 
We use the amplitudes in Eqs.\ (\ref{eq:rho}) and  (\ref{eq:f0}) as the isobar coefficients 
for $B^+ \to \rho^0 \pi^+$ and $B^+\to f_0 \pi^+$, respectively. 
An expression similar to (\ref{isobar}) applies to the CP-conjugate process, 
$B^-\to \pi^-\pi^-\pi^+$, in which the isobar coefficient of $B^- \to \rho^0\pi^-$ 
is given by (\ref{eq:rhob}) while that of $B^- \to f_0 \pi^-$ remains the same as
 (\ref{eq:f0}). In our analysis, we use $\gamma = 65^\circ$ while the relative strong 
 phase $\de_{f_0}$ between the $\rho^0$ and $f_0$ modes is an unknown parameter.
%
\begin{figure}[h]
\begin{center}
\includegraphics[width=0.7\textwidth]{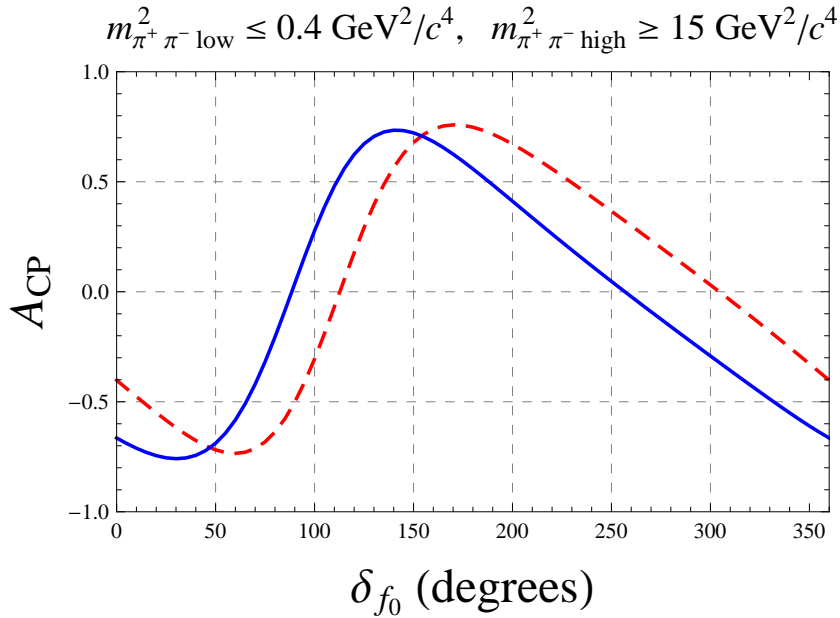}
\end{center}
\caption{Dependence of $A_{CP}(B^+ \to \pi^+ \pi^+ \pi^0)$ with $m^2_{\pi^+
\pi^-\,{\rm low}} < 0.4~{\rm GeV}^2/c^4$ and $m^2_{\pi^+\pi^-\,{\rm high}}>
15~{\rm GeV}^2/c^4$ on the strong phase $\de_{f_0}$, for $\de_V =
- 18^\circ$ (solid) or $18^\circ$ (dashed).
\label{fig:ACPGT}}
\end{figure}

\begin{figure}
\begin{center}
\includegraphics[width=0.7\textwidth]{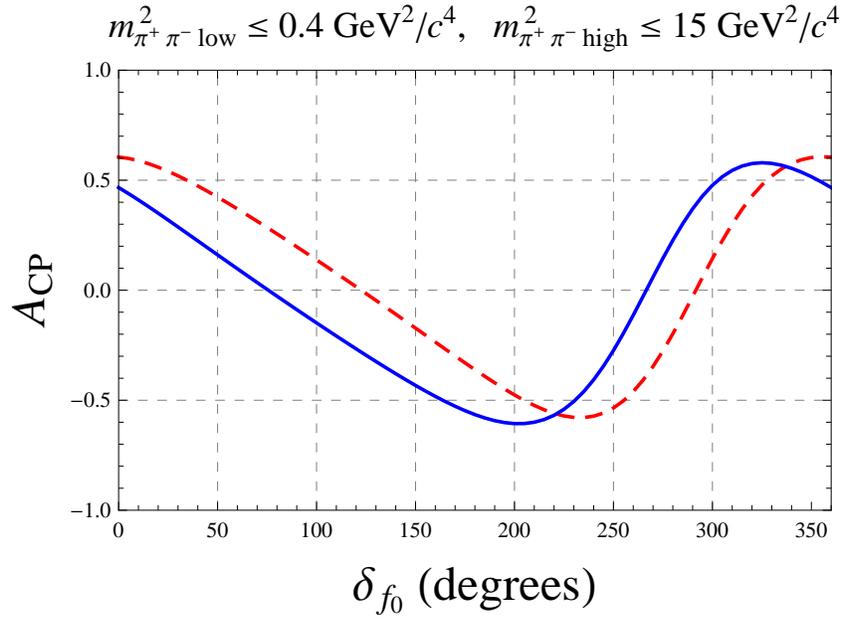}
\end{center}
\caption{Same as Fig.\ \ref{fig:ACPGT} but for $m^2_{\pi^+\pi^-\,{\rm high}} <
15~{\rm GeV}^2/c^4$ 
\label{fig:ACPLT}}
\end{figure}

In Fig.\ \ref{fig:ACPGT} we plot $A_{CP}$ for the Dalitz plot region $m^2_{\pi^+\pi^-\,{\rm low}} <
0.4~{\rm GeV}^2/c^4$ and $m^2_{\pi^+\pi^-\,{\rm high}}$ $> 15~{\rm GeV}^2/c^4$
as a function of $\de_{f_0}$ for $\de_V = - 18^\circ$
as found in the fit of Ref.\ \cite{Chiang:2003pm} (solid) or for $\de_V = 
18^\circ$, as suggested by the discussion below Eq.\ (\ref{ACPrho}) (dashed).
For $\de_V = -18^\circ$, a maximum CP asymmetry of nearly 0.75 is found for
$\de_{f_0} \simeq 140^\circ$, with $A_{CP}$ exceeding 0.5 for a $75^\circ$
range of $\de_{f_0}$ about this value.  The main effect of the reversal of the
sign of $\de_V$ is to shift the plot along the $\de_{f_0}$ axis by an amount
$2\de_V$.
As a consistency check, one can predict the dependence on $\de_{f_0}$ of the
CP asymmetry with $m^2_{\pi^+ \pi^-\,{\rm low}} < 0.4~{\rm GeV}^2/c^4$ and
$m^2_{\pi^+\pi^-\,{\rm high}} < 15~{\rm GeV}^2/c^4$.  The
result is shown in Fig.\ \ref{fig:ACPLT}.  The measured CP asymmetries
for both ranges of $m^2_{\pi^+\pi^-\,{\rm high}}$ should be consistent with
a single value of $\de_{f_0}$.

We have checked that Figs.\,\ref{fig:ACPGT} and \ref{fig:ACPLT} are not affected 
significantly by using a simple pole form for the $f_0(500)$ resonance, or by 
changing its mass and width parameters. We have also studied the effect 
of a decay amplitude to $f_0(980)\pi^+$ by varying its strength and strong phase 
relative to the amplitude for $f_0(500)\pi^+$. The effect on the 
asymmetry for $m^2_{\pi^+\pi^-,\,{\rm low}} < 0.4\,{\rm GeV}^2/c^4$ was found 
to be negligible due to the small overlap of the $f_0(980)$ resonance tail with 
this low invariant mass.
%
\begin{figure}
\begin{center}
\includegraphics[width=0.7\textwidth]{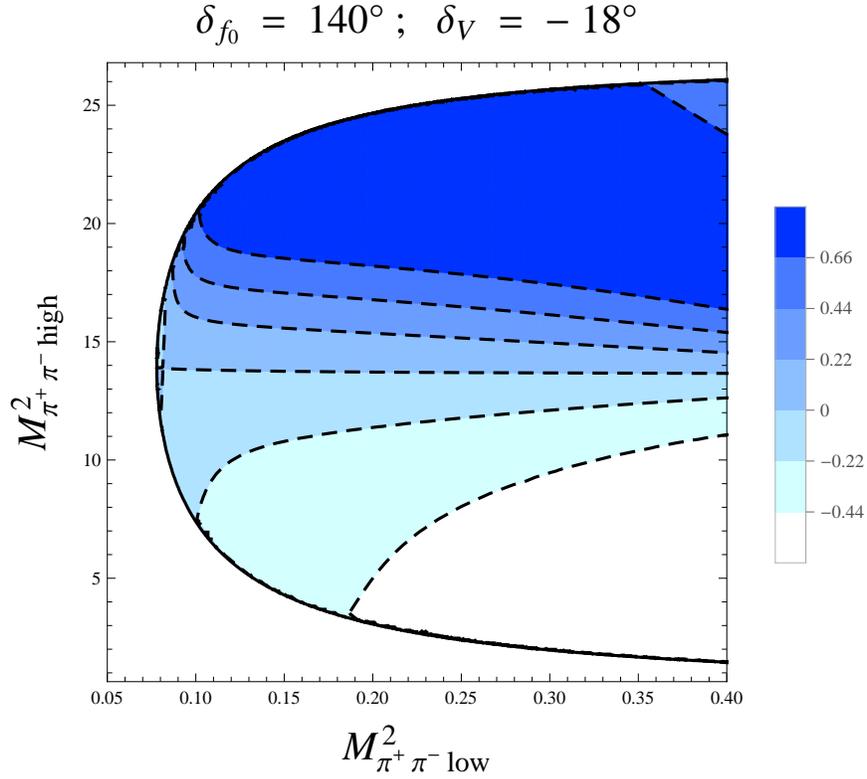}
\end{center}
\caption{CP asymmetry $A_{CP}(B^+ \to \pi^+ (\pi^+\pi^-)_{\rm low~m})$ for
restricted region of the Dalitz plot, with $\de_V = -18^\circ$ and
$\de_{f_0} = 140^\circ$.
\label{fig:ACPdal}}
\end{figure}

As we are introducing only the $\rho \pi$ and $f_0(500) \pi$ final states, we should
expect to reproduce only the region of the Dalitz plot with low
$m^2_{\rm low}$. The values of $A_{CP}(B^+ \to \pi^+ (\pi^+\pi^-)_{\rm low~m})$
are shown in Fig.\ \ref{fig:ACPdal} for 
$m^2_{\rm low} \le 0.4\,{\rm GeV}^2/c^4$, with the choice of parameters 
$\de_V = -18^\circ$ and $\de_{f_0} = 140^\circ$. 
For this specific choice, the CP asymmetry for very low $m^2_{\rm low}$ is 
strongly positive for $m^2_{\rm high} > 15$ GeV/$c^2$ and 
mostly negative for $m^2_{\rm high} < 15$ GeV/$c^2$.

\subsection{$B^\pm \to \pi^\pm K^+ K^-$}

For the $\Delta S = 0$ decays $B^\pm \to \pi^\pm K^+ K^-$, the $B^+$ decay
shows a prominent feature in $m^2_{K^+K^-}$ between threshold and 1.5
GeV$^2/c^4$.  This feature is greatly suppressed in the $B^-$ decay,
accounting for the large CP asymmetry in the second of Eqs.\ (\ref{local}).
The high $K^+K^-$ threshold means that there can be no counterpart of the
(on-shell) $\rho^0$--S-wave interference accounting for the CP asymmetry in
$B^\pm \to \pi^\pm \pi^+ \pi^-$, and information on P-wave $K^+K^-$ resonant
amplitudes above 1 GeV/$c^2$ is fragmentary.
The strong phases in the case of $B^\pm \to \pi^\pm K^+ K^-$ can be completely
distinct from those in $B^\pm \to \pi^\pm \pi^+ \pi^-$, and seem so.

\subsection{$B^\pm \to K^\pm \pi^+\pi^-$}

The Dalitz plot of $B^+ \to K^+ \pi^+\pi^-$ for the low-mass region, 
$m^2(\pi^+\pi^-) < 0.66\,{\rm GeV}^2/c^4$, contains a $\rho^0$ band which 
involves a CP asymmetry~\cite{HFAG}~$A_{CP}(B^+ \to K^+ \rho^0) = 
0.37 \pm 0.11$. Other contributions observed in this low-mass region are a
nonresonant term and contributions from $K^+f_0(980)$ and 
$K^+f_2(1270)$~\cite{Aubert:2008bj,Garmash:2005rv}. A calculation of 
the asymmetry for this restricted region, which may account for the large 
positive asymmetry in the first of Eqs.\ (\ref{local1}), is beyond the scope 
of this Letter.

\subsection{CPT argument}

CPT implies equal partial decay widths for a particle and its antiparticle for 
a closed set of final states connected among themselves by final state 
interactions~\cite{Branco:1999fs}. A simple observation may explain the 
opposite signs of the asymmetries in (\ref{local}) if the decays $B^\pm
\to \pi^\pm X$ for low $M(X)$ are saturated by $X = \pi^+ \pi^-, K^+ K^-$.
Assuming that the CPT theorem holds locally 
where rescattering occurs only between these states, 
an asymmetry in the first
of Eqs.\ (\ref{local}) must be compensated by an asymmetry in the second.
This ignores the effects of neutral pairs in $X$ and of possible
multi-particle realizations of $X$, but it serves at least as a qualitative
guide.

The low-mass $\pi^+ \pi^-$ and $K^+ K^-$ channels are
relatively self-contained, with the only important rescatterings involving
charge exchange and $\pi\pi \leftrightarrow K \bar K$. Rescattering to
multi-particle final states occurs only at typical subenergies $> 1.6$ GeV
\cite{Hyams:1973zf,Grayer:1974cr,Cohen:1980cq,Aston:1987ir}.
The processes giving rise to pairs of final-state neutral particles may be
amenable to evaluation using chiral perturbation theory~\cite{Magalhaes:2011sh}.
Thus the asymmetries in $B^\pm \to \pi^\pm \pi^+ \pi^-$ and $B^\pm \to \pi^\pm
K^+ K^-$ Dalitz plots with low $M(\pi^+\pi^-)$ and $M(K^+ K^-)$, respectively,
could be related to one another through rescattering.
The presence of symmetry breaking leads to different thresholds for $\pi
\pi$ and $K \bar K$ pairs, and imposing different cutoffs on their invariant 
mass is expected to affect this relation. 
This is demonstrated by the asymmetries measured in $B^\pm \to K^\pm X$ 
given in Eq.\,(\ref{local1}).

\section{Conclusion\label{sec:Conc}}

We have examined the CP asymmetries in three-body decays of $B^\pm$ mesons to
charged pions and kaons.  Predictions of ratios of asymmetries on the basis
of U-spin are seen to be obeyed qualitatively, with violations ascribable
to resonant substructure differing for $\pi^+ \pi^-$ and $K^+ K^-$ substates.
Larger CP asymmetries for regions of the Dalitz plot involving low effective
mass of these substates can be undertood qualitatively in terms of large
final-state strong phases; the weak phases are conducive to such large
asymmetries, being nearly maximal.
We conclude that further resolution of this problem must rely either on a
deeper understanding of the resonant substructure in $B \to PPP$ decays,
or further understanding of the hadronization process independently of
resonances.
We have argued that the approximately equal magnitudes and opposite signs
measured 
for asymmetries in $B^+ \to \pi^+\pi^+\pi^-$ and $B^+\to K^+\pi^+\pi^-$ 
may follow from the closure of low-mass $\pi^+\pi^-$ and $K^+K^-$ 
channels involving only $\pi\pi \leftrightarrow K\bar K$ rescattering.

\section*{Acknowledgments}

We thank Ignacio Bediaga and Alberto dos Reis for helpful 
discussions  and for pointing out the work of Ref.\ \cite{Magalhaes:2011sh}.
The work of J. L. R. is supported in part by the United States Department of
Energy under Grant No.\ DE-FG02-90ER40560. The work of B. B. is supported by
NSERC of Canada. B. B. thanks the University of Chicago Theory Group for
hospitality during part of this work.


\begin{thebibliography}{99}
\bibitem{Gronau:1990ka}
  M.~Gronau and D.~London,
  Phys.\ Rev.\ Lett.\  {\bf 65} (1990) 3381.

\bibitem{CKMfitter} J. Charles {\it et al.} (CKMfitter Collaboration), Eur.\
Phys.\ J. C {\bf 41} (2005) 1, periodic updates at
{\tt http://ckmfitter.in2p3.fr/}.

\bibitem{GLW} M. Gronau and D. London, Phys.\ Lett.\ B {\bf 253} (1991) 483;
M. Gronau and D. Wyler, Phys.\ Lett.\ B {\bf 265} (1991) 172;
D. Atwood, I. Dunietz, and A. Soni, Phys.\ Rev.\ Lett.\ {\bf 78} (1997) 3257;
A. Giri, Y. Grossman, A. Soffer, and J. Zupan, Phys.\ Rev.\ D {\bf 68} (2003)
054018.

\bibitem{HFAG} Y. Amhis {\it et al.} (Heavy Flavor Averaging Group),
arXiv:1207.1158, periodic updates at
{\tt http://www.slac.stanford/edu/xorg/hfag}.

\bibitem{Gronau:2005kz}
  M.~Gronau,
  Phys.\ Lett.\ B {\bf 627} (2005) 82 [hep-ph/0508047].

\bibitem{Lin:2008zzaa}
  S.~W.~Lin {\it et al.}  (Belle Collaboration),
  Nature {\bf 452} (2008) 332.
  
  \bibitem{Gronau:1998ep} An approximate equality between the two 
  asymmetries was first proposed in
  M.~Gronau and J.~L.~Rosner,
  Phys.\ Rev.\ D {\bf 59} (1999) 113002
  [hep-ph/9809384]. See however M.~Gronau and J.~L.~Rosner,
  Phys.\ Lett.\ B {\bf 644} (2007) 237
  [hep-ph/0610227].
  
  \bibitem{QCD} For QCD-based calculations of the two asymmetries, equal within
theoretical uncertainties, see
  Y.~Y.~Keum and A.~I.~Sanda,
  Phys.\ Rev.\ D {\bf 67} (2003) 054009 [hep-ph/0209014];
  M.~Beneke and M.~Neubert,
  Nucl.\ Phys.\ B {\bf 675} (2003) 333 [hep-ph/0308039].
 
\bibitem{Grossman:2003qp}
  Y.~Grossman, Z.~Ligeti, Y.~Nir, and H.~Quinn
  Phys.\ Rev.\ D {\bf 68} (2003) 015004 [hep-ph/0303171].

\bibitem{Gronau:2003ep}
  M.~Gronau and J.~L.~Rosner,
  Phys.\ Lett.\ B {\bf 564} (2003) 90 [hep-ph/0304178].
  
\bibitem{Aubert:2008bj}
B.~Aubert {\it et al.}~(BaBar Collaboration),
  Phys.\ Rev.\ D {\bf 78} (2008) 012004
  [arXiv:0803.4451 [hep-ex]].

\bibitem{Garmash:2005rv}
A.~Garmash {\it et al.}~(Belle Collaboration),
Phys.\ Rev.\ Lett.\ {\bf 96} (2006)~251803
  [hep-ex/0512066].

\bibitem{Aubert:2006nu}
B.~Aubert {\it et al.}~(BaBar Collaboration),
  Phys.\ Rev.\ D {\bf 74} (2006) 032003
  [hep-ex/0605003].

\bibitem{Lees:2012kxa}
J.~P.~Lees {\it et al.}~(BaBar Collaboration),
Phys.\ Rev.\ D {\bf 85} (2012) 112010
  [arXiv:1201.5897 [hep-ex]].

\bibitem{Garmash:2004wa}
A.~Garmash {\it et al.}~(Belle Collaboration),
Phys.\ Rev.\ D {\bf 71} (2005) 092003
 [hep-ex/0412066].

\bibitem{Lees:2013kua}
  J.~P.~Lees {\it et al.} (BaBar Collaboration),
  arXiv:1305.4218 [hep-ex].
  
  \bibitem{Aubert:2005sk}
  B.~Aubert {\it et al.}  (BaBar Collaboration),
  Phys.\ Rev.\ D {\bf 72} (2005) 052002
  [hep-ex/0507025].

\bibitem{Aubert:2009av}
B.~Aubert {\it et al.}~(BaBar Collaboration),
Phys.\ Rev.\ D {\bf 79} (2009) 072006
 [arXiv:0902.2051 [hep-ex]].

\bibitem{Aubert:2007xb}
B.~Aubert {\it et al.}~(BaBar Collaboration),
Phys.\ Rev.\ Lett.\ {\bf 99} (2007) 221801
[arXiv:0708.0376 [hep-ex]].

\bibitem{Th-B3P}
 M.~Gronau and J.~L.~Rosner,
 Phys.\ Rev.\ D {\bf 72} (2005) 094031
 [hep-ph/0509155];
 H.~-Y.~Cheng, C.~-K.~Chua and A.~Soni,
 Phys.\ Rev.\ D {\bf 76} (2007) 094006
 [arXiv:0704.1049 [hep-ph]];
 I.~Bediaga, D.~R.~Boito, G.~Guerrer, F.~S.~Navarra and M.~Nielsen,
Phys.\ Lett.\ B {\bf 665} (2008) 30 [arXiv:0709.0075 [hep-ph]];
B.~El-Bennich, A.~Furman, R.~Kaminski, L.~Lesniak, B.~Loiseau and B.~Moussallam,
 Phys.\ Rev.\ D {\bf 79} (2009) 094005
  [Erratum-ibid.\ D {\bf 83} (2011) 039903]
 [arXiv:0902.3645 [hep-ph]];
N.~R.-L.~Lorier, M.~Imbeault and D.~London,  
Phys.\ Rev.\ D {\bf 84} (2011) 034040
[arXiv:1011.4972 [hep-ph]];
 A.~Furman, R.~Kaminski, L.~Lesniak and P.~Zenczykowski,
 Phys.\ Lett.\ B {\bf 699} (2011) 102
 [arXiv:1101.4126 [hep-ph]].

 \bibitem{Aaij:2013sfa}
  R. Aaij {\it et al.}  (LHCb Collaboration),
  arXiv:1306.1246 [hep-ex].

\bibitem{LHCb3P0} R. Aaij {\it et al.}, (LHCb Collaboration), Report No.\
LHCb-CONF-2012-028, presented at the International Conference on High Energy
Physics, Melbourne, Australia, July 2012.

\bibitem{Gronau:2000zy}
 M.~Gronau,
  Phys.\ Lett.\ B {\bf 492} (2000) 297 [hep-ph/0008292].

  \bibitem{He:1998rq}
  X.~-G.~He,
  Eur.\ Phys.\ J.\ C {\bf 9} (1999) 443
  [hep-ph/9810397].

\bibitem{Fleischer:1999pa}
  R.~Fleischer,
  Phys.\ Lett.\ B {\bf 459} (1999) 306 [hep-ph/9903456].

\bibitem{Gronau:2000md}
  M.~Gronau and J.~L.~Rosner,
  Phys.\ Lett.\ B {\bf 482} (2000) 71 [hep-ph/0003119].

\bibitem{Buchalla:1995vs}
  G.~Buchalla, A.~J.~Buras and M.~E.~Lautenbacher,
  Rev.\ Mod.\ Phys.\ {\bf 68} (1996) 1125 [hep-ph/9512380].

\bibitem{Jarlskog:1985ht}
  C.~Jarlskog,
  Phys.\ Rev.\ Lett.\ {\bf 55} (1985) 1039;
I.~Dunietz, O.~W.~Greenberg and D.~-d.~Wu,
 {\it ibid.} {\bf 55} (1985) 2935.
 
 \bibitem{Aaij:2013iua}
  R. Aaij {\it et al.}  (LHCb Collaboration),
  Phys.\ Rev.\ Lett.\  {\bf 110} (2013) 221601
  [arXiv:1304.6173 [hep-ex]].
 
  \bibitem{Gronau:1995hm}
  M.~Gronau, O.~F.~Hernandez, D.~London and J.~L.~Rosner,
  Phys.\ Rev.\ D {\bf 52} (1995) 6356
  [hep-ph/9504326].
  
  \bibitem{Nagashima:2007qn} 
  M.~Nagashima, A.~Szynkman and D.~London,
  Mod.\ Phys.\ Lett.\ A {\bf 23} (2008) 1175 
  [hep-ph/0701199].
  
  \bibitem{Imbeault:2011jz}
  M.~Imbeault and D.~London,
  Phys.\ Rev.\ D {\bf 84} (2011) 056002
  [arXiv:1106.2511 [hep-ph]].
  
  \bibitem{He:2013vta}
  X.~-G.~He, S.~-F.~Li and H.~-H.~Lin,
  JHEP {\bf 1308} (2013) 065
  [arXiv:1306.2658 [hep-ph]].
  
  \bibitem{Gronau:2013mda}
  M.~Gronau,
  arXiv:1308.3448 [hep-ph].

\bibitem{Zhang:2013oqa}
  Z.~-H.~Zhang, X.~-H.~Guo and Y.~-D.~Yang,
  Phys.\ Rev.\ D {\bf 87} (2013) 076007 [arXiv:1303.3676 [hep-ph]].

\bibitem{Cheng:2013fba}
  H.~-Y.~Cheng, C.~-K.~Chua, K.~-C.~Yang and Z.~-Q.~Zhang,
  Phys.\ Rev.\ D {\bf 87} (2013) 114001  
  [arXiv:1303.4403 [hep-ph]].

\bibitem{Chiang:2003pm}
  C.~-W.~Chiang, M.~Gronau, Z.~Luo, J.~L.~Rosner and D.~A.~Suprun,
  Phys.\ Rev.\ D {\bf 69} (2004) 034001 [hep-ph/0307395].

\bibitem{Lipkin:1980tk}
  H.~J.~Lipkin,
  Phys.\ Rev.\ Lett.\ {\bf 46} (1981) 1307;
  Phys.\ Lett.\ B {\bf 254} (1991) 247; {\bf 415} (1997) 186;
  {\bf 433} (1998) 117.
  
  \bibitem{Branco:1999fs} 
  G.~C.~Branco, L.~Lavoura and J.~P.~Silva, 
  CP violation, Int.\ Ser.\ Monogr.\ Phys.\  vol. 103,
  Oxford Science Publications,  1999, page 58.

\bibitem{Hyams:1973zf}
  B.~Hyams,
 {\it et al.},
  Nucl.\ Phys.\ B {\bf 64} (1973) 134
   [AIP Conf.\ Proc.\  {\bf 13} (1973) 206].

\bibitem{Grayer:1974cr}
  G.~Grayer,
{\it et al.},
  Nucl.\ Phys.\ B {\bf 75} (1974) 189.

\bibitem{Cohen:1980cq}
  D.~H.~Cohen,
{\it et al.},
  Phys.\ Rev.\ D {\bf 22} (1980) 2595.

\bibitem{Aston:1987ir}
  D.~Aston, N.~Awaji, T.~Bienz, F.~Bird, J.~D'Amore, W.~M.~Dunwoodie,
  R.~Endorf, and K.~Fujii {\it et al.},
  Nucl.\ Phys.\ B {\bf 296} (1988) 493.

\bibitem{Magalhaes:2011sh}
  P.~C.~Magalhaes, M.~R.~Robilotta, K.~S.~F.~F.~Guimaraes, T.~Frederico, W.~de Paula, I.~Bediaga, A.~C.~d.~Reis and C.~M.~Maekawa {\it et al.},
  Phys.\ Rev.\ D {\bf 84} (2011) 094001
  [arXiv:1105.5120 [hep-ph]].

\end{thebibliography}
\end{document}